\documentclass[preprintnumbers,aps,pra,twocolumn,showpacs]{revtex4}

\usepackage{graphicx}
\usepackage{dcolumn} 
\usepackage{amsmath}
\usepackage{exscale}
\usepackage{bm}      
\usepackage{color}
\usepackage{epsfig,psfrag,subfigure,subeqnarray}
\usepackage{bbm}
\usepackage{amsbsy}
\usepackage{amssymb}
\usepackage{enumitem}

\def\lan{\langle}
\def\ran{\rangle}

\def\vk{{\bf k}}

\def\vQ{{\bf Q}}

\def\vp{{\bf p}}
\def\vR{{\bf R}}

\newcommand{\bd}{\begin{equation}}
\newcommand{\ed}{\end{equation}}
\newcommand{\be}{\begin{equation}}
\newcommand{\ee}{\end{equation}}
\newcommand{\bt}{\begin{split}}
\newcommand{\et}{\end{split}}

\newcommand{\bn}{\begin{align}}
\newcommand{\en}{\end{align}}

\newcommand{\bea}{\begin{eqnarray}}
\newcommand{\eea}{\end{eqnarray}}
\newcommand{\ba}{\begin{array}}
\newcommand{\ea}{\end{array}}
\newcommand{\nn}{\nonumber}

\begin{document}

\title{Effect of Pauli blocking on coherent states of composite bosons}

\author{Shiue-Yuan Shiau$^1$}
\email{shiau.sean@gmail.com}
\author{Monique Combescot$^2$}
\affiliation{$^1$ Department of Physics, National Cheng Kung University, Tainan, 701 Taiwan}
\affiliation{$^2$ Institut des NanoSciences de Paris, Universit\'e Pierre et Marie Curie, CNRS, 4 place Jussieu, 75005 Paris}

\begin{abstract}
 The quantum nature of elementary bosons can be completely erased by using coherent states known as Glauber states. Here, we consider composite bosons (cobosons) made of two fermions and look for the possibility to erase  the bosonic quantum nature of the field \textit{and} the fermionic quantum nature of its constituents, when the distribution of the coboson Schmidt decomposition is either flat like Frenkel excitons, or localized like Wannier excitons. We show that for Frenkel-like cobosons, complete erasure of the field quantum nature is possible up to a density which corresponds to a sizable fraction of the number of fermion-pair states making the coboson at hand. At higher density, the Pauli exclusion principle between fermionic constituents shows up dramatically. It induces: (i) the decrease of number-operator eigenvalues down to 1 and the increase of the number-state second-order correlation function up to 2; (ii) the disappearance of the usual sharp peak in the coherent-state distribution and the increase of the coherent-state second-order correlation function up to 2. It also is possible to construct number states and coherent states for Wannier-like cobosons, but their forms are far more complex. Finally, we show that Pauli blocking makes the coboson coherent states qualitatively different from the Anderson's ansatz.
\end{abstract}

\date{\today }

\maketitle

\section{Introduction}

 The milestone Hanbury-Brown-Twiss optical interferometry experiments\cite{HBT1,HBT2}, made in 1956, have opened the route to studying optical quantum regime through direct measurement of high-order correlation functions for photons. While the first-order correlation function probes phase coherence as it is a one-body quantity, the photon quantum nature and the coherence properties of the field show up in higher-order correlation functions.  \

Shortly after Hanbury-Brown-Twiss experiments, Glauber\cite{Glauber1963}  laid the groundwork for a modern theory of quantum optical coherence.
In particular, he showed that it is possible to construct a linear combination of elementary-boson Fock (or number) states that gives all $n$-order correlation functions exactly equal to 1. Such a state is commonly called coherent or Glauber state. This state is classical in the sense that the uncertainty product $\Delta p\Delta x$ calculated within this state is equal to its minimum value, $\hbar/2$. Glauber states are physically important because they describe output light sources of single-mode lasers\cite{Yarivbook,Loudon}.\

In recent years, measurements of high-order correlations have been extended to massive  bosons such as cold bosonic atoms or cold fermionic-atom dimers\cite{Yasuda1997,Hodgman2011,Guarrera2011,Manning2013}. Dall et all\cite{Dall2013NP} have performed an impressive feat by measuring atom-atom correlation functions for ultracold $^4$He up to six order. Mature optical calibration and precise control of trapped cold-atom gases have opened a new area of physics with precise studies of quantum collective properties for massive bosonic particles, as in Bose-Einstein condensate.\

Composite bosons (cobosons for short) behave as elementary bosons when they are tightly bound; coherence measurements then agree with Glauber's predictions. By contrast, for loosely-bound cobosons  such as semiconductor excitons, the underlying  fermionic constituents play a crucial role in the coherence properties of these particles, through Pauli blocking induced by the Pauli exclusion principle. We have recently shown\cite{MoniquePRA2009} that the second-order correlation function of $N$ identical cobosons, each made of two fermions, is larger than that for $N$ identical elementary bosons. Corrections coming from fermion exchange between cobosons depend on the dimensionless parameter that controls coboson many-body effects, namely $\eta \equiv N(a / L)^D$, where $D$ is the space dimension, $L$ the sample size, and $a$ the Bohr radius of the two fermions making the coboson. As precise studies of the coherence properties for composite bosons require to take into account the Pauli exclusion principle, there is an urgent need to include its consequences in Glauber's theory.\

In this work, we  address the following important question: is it possible to construct Glauber states from cobosons made of two fermions, that is, eigenstates of the coboson destruction operator, such that all correlation functions $g_n$ exactly equal 1? To put it differently, can the bosonic quantum nature of the coboson field \textit{and} the fermionic nature of its  constituents be erased completely? 

Glauber showed that the answer is yes for elementary bosons. This  is not so much a surprise because elementary-boson Fock states from which Glauber states are constructed have correlation functions smaller than 1. The construction of Glauber states from cobosons faces many more difficulties. First, $g_2$ for coboson Fock states is larger than its counterpart for elementary bosons\cite{MoniquePRA2009}. Secondly, even if we neglect fermion-fermion interaction, the underlying fermionic nature of cobosons is very strong: when the coboson density becomes sizable, coboson Fock states are so altered by  the Pauli exclusion principle\cite{PogosovJETP2010}  that their bosonic properties completely disappear.  Despite these difficulties, we will show that it is possible to construct Glauber states for cobosons with a second-order correlation function equal to 1, not only in the expected very dilute limit, but also in a dense regime where Pauli blocking comes into play.\

We  approach this problem through the coboson many-body formalism\cite{MoniqPhysreport,Moniquebook}. One merit of this formalism is to transparently reveal the consequences of  fermion exchanges between cobosons. We consider cobosons in their  Schmidt decomposition (see Eq.~(\ref{Bdagvarphip})), with a $|\varphi_p|$ distribution either flat, that is, completely delocalized like Frenkel excitons\cite{Agranovichbook}, or localized like Wannier excitons\cite{Moniquebook}. In this work, we focus on the effect of the Pauli exclusion principle between the coboson constituents. Their fermionic  quantum nature affects the elementary boson results in different ways, but in both cases, we succeed to construct Fock  and Glauber states.  \

We show that for Frenkel-like cobosons with a flat $|\varphi_p|$ distribution, Glauber states maintain a second-order correlation function equal to 1 up to a coboson number as large as a sizable fraction ($\simeq 16\%$) of the number of pair states making them. For Wannier-like cobosons having a localized $\varphi_p$ distribution --- a situation definitely more complex than when the distribution is flat --- we provide a formal procedure to derive Fock states and Glauber states. Finally, we  show that, except in the very dilute limit, Glauber states for Frenkel-like and Wannier-like cobosons qualitatively differ from the ansatz state $e^{\alpha B^\dag}|0\ran$ commonly used to study coboson condensation and coherence, as first proposed by Anderson\cite{Anderson1958}. \

This paper is organized as follows: In Sec.~\ref{elemboson}, we briefly introduce Fock states and Glauber states for elementary bosons with creation operator $A^\dag$. These states are, respectively, defined as eigenstates of the number operator $A^\dag A$ and the destruction operator $A$.  In Sec.~\ref{sec:cb}, we first connect the usual decompositions of Frenkel and Wannier excitons to the Schmidt decomposition of composite bosons commonly used in quantum information. We then present the two key commutators of the coboson many-body formalism that we use to calculate many-body effects induced by fermion exchange between cobosons. We end this section by briefly discussing what we have called ``moth-eaten effect" which occurs in all coboson systems and which comes from the Pauli exclusion principle between coboson fermionic constituents. In Sec.~\ref{ssec3}, we focus on Frenkel-like cobosons whose $\varphi_p$ distribution is just a phase, and construct their Fock and Glauber states. We study the probability distribution and second-order correlation function of these states as a function of the number  of fermion-pair states that enter the Frenkel-like coboson operators, and as a function of the average coboson number of the Glauber state at hand.  In Sec.~\ref{ssec4}, we turn to Wannier-like cobosons whose Schmidt distribution is localized. Although the procedure is far more complicated, we show that it is possible to construct Fock states and Glauber states for this coboson field, and we study the dependence of these quite complex states on the coboson field characteristics. In Sec.~\ref{ssec:6}, we compare  Glauber states of the above two types of cobosons with the Anderson's ansatz. In Sec.~\ref{sota}, we discuss previous research on coboson Glauber states. We then conclude.

\section{Elementary bosons\label{elemboson}}
To better grasp the consequences of the coboson composite nature which markedly distinguishes cobosons from elementary bosons, let us briefly recall the results for elementary bosons.\

For a classical field $A(t)$, the second-order correlation function evaluated in the system ensemble,
\be
g_2 (t)\equiv \frac{\langle A^*(t) A^*(t) A(t) A(t) \rangle}{(\langle A^*(t) A(t) \rangle)^2}\, ,
\ee
is exactly equal to 1, while classical noises always yield $g_2(t)$ larger than 1. By contrast, for a quantum field of elementary bosons --- as photon field --- with creation operator $A^\dagger$ such that $\left[A,A^\dagger\right]_-=1$, the $\vert \phi \rangle$ state second-order correlation function,
\be
g_2 = \frac{\langle \phi \vert A^{\dagger 2} A^2 \vert \phi \rangle}{\langle \phi \vert \phi \rangle} \left( \frac{\langle \phi \vert \phi \rangle}{\langle \phi \vert A^\dagger A \vert \phi \rangle} \right)^2\, ,\label{eq:g2}
\ee 
can be less than 1. The lowest possible value for $g_2$, obtained for the Fock state $A^{\dagger N} \vert 0 \rangle$ with $\vert 0 \rangle$ denoting the vacuum, is equal to $ 1 - 1/N$ (see \ref{app:sec1}), which  is the lowest possible $g_2$ value an $N$-elementary boson state can reach. \

It  is possible to erase this bosonic quantum nature and achieve all-order correlation functions equal to 1  through a linear combination of Fock states, as first shown by Glauber\cite{Glauber1963}. This so-called Glauber state, also known as coherent state, is defined in terms of normalized Fock states $\vert N \rangle = (N!)^{-1/2} A^{\dagger N} \vert 0 \rangle$ as 
\be
\vert \phi_\alpha \rangle = \sum_{N=0}^\infty \frac{\alpha^N}{\sqrt{N!}} \vert N \rangle\, .\label{defcoherentstateelemB}
\ee
It is easy to see that $\vert \phi_\alpha \rangle$ is eigenstate of the boson destruction operator $A$ with eigenvalue $\alpha$, 
\be\label{eq:sixprime}
A \vert \phi_\alpha \rangle=\sum_{N=0}^\infty \frac{\alpha^N}{\sqrt{N!}}\sqrt{N} \vert N-1 \rangle = \alpha \vert \phi_\alpha \rangle\, .
\ee

As  Glauber states for elementary bosons have a Poisson distribution over the Fock states $|N\ran$, their distribution as a function of $N$ is peaked at $|\alpha|^2$ (see Fig.~\ref{Fig3}), which is equal to the average boson number, $\langle \phi_\alpha \vert A^\dagger A \vert \phi_\alpha \rangle/\langle \phi_\alpha \vert \phi_\alpha \rangle $. Since $\langle \phi_\alpha \vert A^{\dagger 2} A^2 \vert \phi_\alpha \rangle/ \langle \phi_\alpha \vert \phi_\alpha \rangle$ gives $ | \alpha|^4 $, we readily find that the second-order correlation function $g_2$ for the $\vert \phi_\alpha \rangle$ state is exactly equal to 1 whatever $\alpha$. And similarly for \textit{all} higher-order correlation functions.

\section{Composite bosons made of two fermions\label{sec:cb}}

The quantum nature of composite bosons made of two fermions is far more subtle than that of elementary bosons. The fundamental complexity lies in the impossibility of associating a particular pair of fermions to a coboson. This fermion indistinguishability  allows fermion exchanges between cobosons, these exchanges giving rise to the dimensionless ``Pauli scatterings" of the coboson many-body formalism\cite{Moniquebook}. The Pauli exclusion principle between coboson fermionic constituents induces the so-called ``moth-eaten effect"\cite{PogosovJETP2010} which, in particular, forbids piling up more cobosons than the number of fermion-pair states from which the cobosons are made. The Pauli exclusion principle being fundamentally unquenchable, the induced ``moth-eaten effect" is extremely robust and shows up in all problems involving cobosons. In the following, we will show its dramatic consequences on Fock and Glauber states of  coboson fields.

\subsection{Coboson operators\label{ssec2:a}}

The most general form for creation operators of cobosons made of two fermions depends on two indices,
\be
B^\dag=\sum_n\sum_m f_{nm} a^\dag_n b^\dag_m\, ,\label{eq:Bcreation2indces}
\ee
where the operators $(a^\dag, b^\dag)$ create the two fermions at hand. It is however possible to write this operator in a diagonal form using what fundamentally is a change of basis, known as Schmidt decomposition, namely
\be
B^\dag=\sum_p \varphi_p \alpha^\dag_p \beta^\dag_p\equiv \sum_p \varphi_p B^\dag_p\label{Bdagvarphip}
\ee
with $\alpha^\dag_p=\sum_n \mathcal{A}_{np} a^\dag_n$ and $\beta^\dag_p=\sum_m \mathcal{B}_{pm}b^\dag_m$, the matrices $\mathcal{A}$ and $\mathcal{B}$, with components $\mathcal{A}_{np}$ and $\mathcal{B}_{pm}$, being unitary in order to preserve fermion anticommutation relations. The $f_{nm}$ and $\varphi_p$ prefactors are related by $f_{nm}=\sum_p \mathcal{A}_{np} \varphi_p \mathcal{B}_{pm}$. To bridge the above coboson creation operator commonly used in quantum information to the ones for semiconductor excitons, we first note that the creation operator of a Frenkel exciton made of electron-hole pairs located on lattice sites $\vR_n$ reads\cite{Agranovichbook,MoniquePRB2008} 
\be
 B_\vQ^\dag= \frac{1}{\sqrt{N_s}} \sum_{n=1}^{N_s} e^{i\vQ\cdot \vR_n} a^\dag_n b^\dag_n\, ,\label{eq:Bdagthetap}
\ee 
where $N_s$ is the number of lattice sites. The tight-binding approximation, valid for these excitons, makes their creation operators depend on a single index $n$ only. So, they appear in a  diagonal Schmidt form.\

By contrast, the creation operator\cite{Moniquebook} of a Wannier exciton a priori depends on two quantum indices, $i=(\vQ_i,\nu_i)$,
\be
B^\dag_i= \sum_{\vk_e }\sum_{\vk_h} a^\dag_{\vk_e}b^\dag_{\vk_h} \lan \vk_h, \vk_e|i\ran\, .
\ee
By noting that momentum conservation imposes $\vQ_i=\vk_h +\vk_e$, it is possible to write the $i$ exciton creation operator in a diagonal form, as a linear combination of pairs having different relative-motion momentum $\vp$, 
\be
B^\dag_{\vQ_i \nu_i}= \sum_{\vp} a^\dag_{\vp+\gamma_e \vQ_i}b^\dag_{-\vp +\gamma_h \vQ_i} \lan \vp|\nu_i\ran\, ,
\ee 
with $\gamma_e=1-\gamma_h=m_e/(m_e+m_h)$, in order for the electron and hole kinetic energies, $\vk_e^2/2m_e+\vk_h^2/2m_h$, to split into  a center-of-mass and a relative-motion contribution, $\vQ_i^2/2M+\vp^2/2\mu$,
with $M=m_e+m_h$ and $\mu^{-1}=m_e^{-1}+m_h^{-1}$. The above creation operator has the form of Eq.~(\ref{Bdagvarphip}) for $\alpha^\dag_\vp=a^\dag_{\vp+\gamma_e \vQ_i}$ and $\beta^\dag_\vp=b^\dag_{-\vp +\gamma_h \vQ_i}$.\ 

As we here are  interested in the quantum aspects of cobosons induced by the Pauli exclusion principle, we can stay with the diagonal form (\ref{Bdagvarphip}), which renders the algebra of the coboson formalism far simpler. We are going to successively consider: (i) Frenkel-like cobosons, which are fully delocalized in space, their $\varphi_p$ distribution being just a phase; (ii) Wannier-like cobosons, which have a spatial extension $a$, their $\varphi_p$ distribution being peaked with a momentum extension $\sim 1/a$.

\subsection{Many-body formalism for fermion exchange\label{ssec2:b}}


We here consider coboson operator $B^\dag$ written in a diagonal form (\ref{Bdagvarphip}), and normalized by $\sum_p |\varphi_p|^2=1$. The effects of the coboson composite nature follow from two commutators\cite{MoniqPhysreport,Moniquebook}, 
\bea
\big[B,B^\dag\big]_-&=&1-D\, ,\label{eq:comBB}\\
\big[D,B^\dag\big]_-&=&2\sum_p |\varphi_p|^2 \varphi_p \alpha^\dag_p \beta^\dag_p \equiv C^\dag\, .\label{eq:comDB}
\eea
the operator $D$ being such that $D|0\ran=0$. Iteration of these two commutators give
\bea
\big[D,B^{\dag N}\big]_-&=&NC^\dag B^{\dag N-1}\, ,\\
\big[B,B^{\dag N}\big]_-&=& N B^{\dag N-1} (1-D)-\frac{N(N-1)}{2}C^\dag B^{\dag N-2}\, .\nn\\ \label{eq:comBBN}
\eea

 Using the above equation, we find that $N$-coboson states $|\psi_N\ran=B^{\dag N}|0\ran $ are related by
\be
B|\psi_N\ran=N|\psi_{N-1}\ran-\frac{N(N-1)}{2}C^\dag |\psi_{N-2}\ran\, .\label{eq:BpsiN}
\ee
The $C^\dag$ term results from the coboson composite nature. It makes the $|\psi_N\ran$ normalization factor 
\be
\lan\psi_N|\psi_N\ran=N!F_N\, ,\label{pshiNnormalifact}
\ee
different from the normalization factor $N!$ of the elementary-boson Fock state $A^{\dag N}|0\ran$.
The $F_N$ factor, equal to 1 for elementary bosons, decreases from $F_1=1$ when $N$ increases, because of the moth-eaten effect between $N$ cobosons.\

We wish to mention that $B|\psi_N\ran$ is often divided\cite{Lawpra2005,Leepra2013} into a state along $|\psi_{N-1}\ran$ and a state orthogonal to $|\psi_{N-1}\ran$. One is then tempted to drop this orthogonal part, as done in Ref.~\onlinecite{Leepra2013}, while keeping $F_N/F_{N-1}\not=1$ as the unique signature of the coboson composite nature, which is fully inconsistent (see \ref{app:sec2}).

\subsection{Moth-eaten effect\label{ssec2:c}}

We have often used the vivid word, ``moth-eaten," to describe a many-body effect induced by the Pauli exclusion principle that occurs in \textit{all} systems of composite particles made of fermions. To physically grasp this effect, consider the single coboson state $|\psi_1\ran=B^{\dag}|0\ran $. If  a $B^{\dag}$ coboson is added to this state, one fermion-pair state cannot be used by the second $B^{\dag}$ operator of the $B^{\dag}|\psi_1\ran=|\psi_2\ran$ state because of the Pauli exclusion principle between the fermionic constituents of the two $B^{\dag}$ operators, as if a little moth had eaten one state among the fermion-pair states $p$ that enter the $ \varphi_p$ distribution. In the same way, a $B^{\dag}$ coboson added to the $N$-coboson state $|\psi_N\ran$ has $N$ fermion-pair states missing, as if $N$ little moths had eaten these states from the $ \varphi_p$ distribution.  \

A very striking way in which this moth-eaten effect shows up is through the $N$-coboson normalization factor $N! F_N$, which is markedly different from $N!$ for $N$ elementary bosons. Indeed, the factor $F_N$ decreases exponentially when $N$ increases, from $F_1=1$ down to zero when $N$ reaches the number $N_s$ of fermion-pair states $p$ that enter the $ \varphi_p$ distribution, as in the case of Frenkel-like cobosons. For $N>N_s$, the $|\psi_N\ran$ state reduces to zero, the little moths having eaten all the $N_s$ pair states that constitute the $ \varphi_p$ distribution.

\section{ Composite bosons with a flat distribution\label{ssec3}}
Let us first consider cobosons having constant $ |\varphi_p|$, that is, $\varphi_p$  equal to a phase as for Frenkel excitons. The coboson creation operator then reads
\be
\hat B^\dag= \frac{1}{\sqrt{N_s}} \sum_{p=1}^{N_s} e^{i\theta_p} \alpha^\dag_p \beta^\dag_p\, ,\label{eq:Bdagthetap}
\ee
where $N_s$ is the number of pair states used to make the coboson at hand. Indeed, since $ |\varphi_p|$ is a constant, this number must be finite in order to have $\sum_p |\varphi_p|^2=1$ that makes  $\hat B^\dag$  normalized. \

 It is easy to check that, for $\hat D$ defined as in Eq.~(\ref{eq:comBB}), the $\hat C^\dag$ operator reduces to
\be
\hat C^\dag=\big[\hat D,\hat B^\dag\big]_-=\frac{2}{N_s}\hat B^\dag\, .
\ee
Inserting the above equation into Eq.~(\ref{eq:BpsiN}) we find, for $|\hat\psi_N\ran=\hat B^{\dag N}|0\ran$,
\be
\hat B|\hat\psi_N\ran=N\left(1-\frac{N-1}{N_s}\right)|\hat\psi_{N-1}\ran\, .\label{eq:BhatpsiN}
\ee
So, the state orthogonal to $|\hat\psi_{N-1}\ran$ exactly cancels (see \ref{app:sec2}). \

The normalization factor $\lan \hat\psi_N|\hat\psi_N\ran=N!\hat F_N$ then takes a compact form 
\be
\hat F_N=\left(1-\frac{N-1}{N_s}\right) \cdots \left(1-\frac{1}{N_s}\right)=\frac{N_s!}{N_s^N (N_s-N)!}\label{defFN}
\ee
for $1\leq N\leq N_s$, and $\hat F_N=0$ for $N> N_s$: indeed, because of the Pauli exclusion principle between coboson fermionic constituents, it is not possible to pile up more cobosons than the number of pair states entering the $\hat B^\dag$ operator. The above equation gives $\hat F_1=1$ and $\hat F_{N+1}/\hat F_N=1-N/N_s$. For $N_s$ large, $\hat F_{N_s}\simeq e^{-N_s}$, while for $(N_s,N_s-N)$ both large, $\hat F_N\simeq e^{-N} (1-N/N_s)^{N-N_s}$.  \


\begin{figure}[t!]
\begin{center}
\includegraphics[trim=0.2cm 0cm 0.7cm 0cm,clip,width=3.2in] {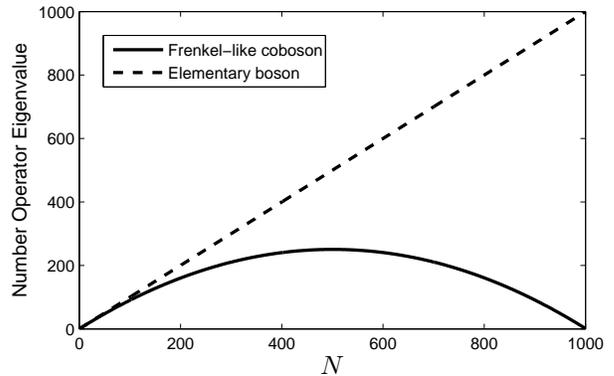}
   \caption{\small Number-operator eigenvalues for elementary bosons, as given in Eq.~(\ref{Nstateeemboso}) and for Frenkel-like cobosons, as given in Eq.~(\ref{eq:numberopereigen}) when $N_s=1000$.  }
   \label{Fig1}
\end{center}
\end{figure}

\subsection{Fock states}

 The normalized $N$-coboson state constructed on the $\hat B^\dag$ operator reads $|\hat N\ran =(N! \hat F_N)^{-1/2}  |\hat\psi_N\ran$.
With the help of  Eq.~(\ref{eq:BhatpsiN}), we  find
\be
\hat B^\dag \hat B| \hat N\ran =N\left(1-\frac{N-1}{N_s}\right)| \hat N\ran\, .\label{eq:numberopereigen}
\ee
So, $|\hat N\ran$ is eigenstate of the number operator $\hat B^\dag \hat B$, but its eigenvalue is decreased from the elementary boson value $N$ down to $N\big(1-(N-1)/N_s\big)$, because of Pauli blocking between the coboson fermionic constituents (see Fig.~\ref{Fig1}). The maximum value this eigenvalue can reach is $(N_s+1)^2/4N_s$ for $N$ equal to $N^*=(N_s+1)/2$, the eigenvalue then being equal to $\simeq N^*/2$. The number-operator eigenvalue keeps deviating from $N$ as $N$ increases, until it reduces to 1 when the  coboson number  reaches the number $N_s$ of pair states making the coboson. This behavior is markedly different from the elementary boson counterpart. \

\begin{figure}[t!]
\begin{center}
\includegraphics[trim=0.5cm 0cm 0.7cm 0.5cm,clip,width=3.1in] {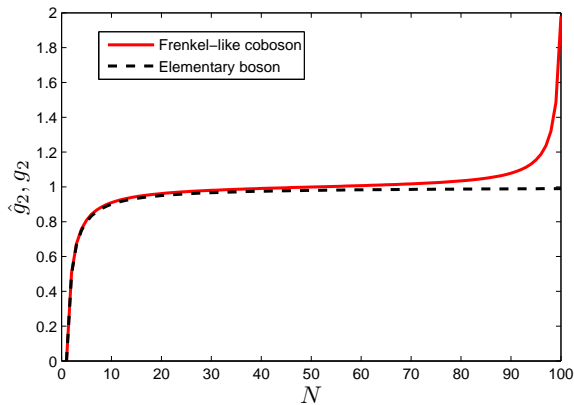}
   \caption{\small (color online) Second-order correlation functions $\hat g_2$ for elementary bosons, as given by $g_2=1-1/N$, and  for Frenkel-like cobosons, as given in Eq.~(\ref{g2frenkelexciton}) when $N_s=100$.  }
   \label{Fig2}
\end{center}
\end{figure}

This eigenvalue decrease has important consequences on the second-order correlation function $\hat g_2$ of the Fock state $| \hat N\ran$. Indeed, as
\be
 \lan\hat N |
\hat B^{\dag2}\hat B^2
  | \hat N\ran =N (N-1)\!\left(1-\frac{N\!-\! 1}{N_s}\right)\!\!\left(1-\frac{N\!-\!2}{N_s}\right)\, ,
\ee
the normalized $| \hat N\ran$ state has a $\hat g_2$ given by
\be
\hat g_2=\frac{ \lan\hat N |
\hat B^{\dag2}\hat B^2
  | \hat N\ran}{ \lan\hat N |
\hat B^{\dag}\hat B
  | \hat N\ran^2}=\left(1-\frac{1}{N}\right)\frac{N_s+2-N}{N_s+1-N}\, .\label{g2frenkelexciton}
\ee

Figure \ref{Fig2} shows the second-order correlation function for Fock states of Frenkel-like cobosons and of elementary bosons. When the coboson number $N$ is much smaller than the number $N_s$ of pair states making the coboson, the $\hat g_2$ value is essentially equal to that of $N$ elementary bosons, as physically reasonable since the effect of Pauli blocking is small. When $N$ reaches $1+N_s/2$, $\hat g_2$ is exactly equal to 1: the bosonic quantum nature of the coboson field starts to disappear. When $N$ approaches $N_s$, $\hat g_2$ increases rapidly until it reaches $2(1-1/N_s)$ for $N=N_s$. The effect of Pauli blocking becomes noticeable when the coboson number $N$ is a sizable fraction of the number $N_s$ of pair states making the coboson.  \

\subsection{Quasi-Glauber states}

Let us now look for Glauber states, that is, eigenstates of the coboson destruction operator $\hat B$, as a linear combination of coboson Fock states,
\be
|\hat \phi_\alpha\ran=\sum_{N=0}^{N_s}  x_N |\hat N\ran\, .\label{eq:hatphialpha01}
\ee 
Compared with  Glauber states for elementary bosons where there is no upper limit for their Fock states $|N\ran$, the fact that $|\hat N\ran$ states only exist for $N\leq N_s$, because of Pauli blocking, has a dramatic consequence on the possible construction of $\hat B$ eigenstates. Indeed, 
 \be
\hat B |\hat \phi_\alpha\ran=\sum_{N=0}^{N_s-1} x_{N+1} \sqrt{(N+1)\left(1-\frac{N}{N_s}\right)} |\hat N\ran\, ,
\ee
has an upper $N$ limit equal to $N_s-1$ instead of $N_s$ as for $|\hat \phi_\alpha\ran$; so, the best we can do is to find a ``quasi-Glauber state" $|\hat \phi_\alpha\ran$ such that
 \be
 \hat B |\hat \phi_\alpha\ran=\alpha\Big(|\hat \phi_\alpha\ran- x_{N_s} |\hat N_s\ran\Big)\, , \label{eq:Bhatpsi_alpha}
  \ee
  and to hope that the missing $\alpha  x_{N_s} |\hat N_s\ran$ state has no sizable consequence. From the above equation, we find that the $ x_N$'s for $N< N_s$ are  related through 
 $\alpha  x_{N-1}= x_N \sqrt{N(N_s-N+1)/N_s}$. For $x_0$ taken equal to 1, this gives
\be
 x_N=\frac{\alpha^N}{\sqrt{N!\hat F_N}}\label{defhatxN}
\ee
with $\hat F_0=1$ by convention.\

 Compared to elementary bosons, the fermionic nature of the $\hat B^\dag$ constituents shows up in two ways:
 
 $\bullet$ through an increase of the $| x_N|^2$ distribution as compared with the Poisson distribution $|\alpha|^{2N}/N!$ for elementary bosons given in Eq.~(\ref{defcoherentstateelemB}), since $\hat F_N$ is a monotonously decreasing function of $N$, starting from  $\hat F_1=1$. Let us consider the two extreme cases: $N=1$ and $N=N_s$. As Pauli blocking starts to appear for two cobosons, we have $| x_1|^2=|\alpha|^2$ whatever $N_s$. By contrast, $\hat F_{N_s}$ for $N_s \gg 1$ is exponentially small, $\hat F_{N_s}\simeq e^{-N_s}$; so, we get from Eq.~(\ref{defhatxN}) that
\be
|x_{N_s}|^2=\frac{|\alpha|^{2N_s}}{N_s!\hat F_{N_s}}\simeq \left(\frac{e^2 |\alpha|^2}{N_s}\right)^{N_s}\, .
\ee
We find that $|x_{N_s}|^2$, equal to $e^{2|\alpha|^2}$ for $N_s=|\alpha|^2$, reaches a maximum value $e^{e|\alpha|^2}$ for $N_s=e|\alpha|^2$, and then decreases down to 1 for $N_s=e^2|\alpha|^2$. As  shown below, this rapid increase of  $| x_{N_s}|^2$  from 1 to $e^{e|\alpha|^2}$ as $|\alpha|^2$ increases from $N_s/e^2$ to $N_s/e$ dramatically changes the shape of the $| x_{N}|^2$ distribution. \

 
 $\bullet$ through the fact that $|\hat \phi_\alpha\ran$ cannot be \textit{exact} eigenstate of the destruction operator $\hat B$. This substantially changes the second-order correlation function of the $|\hat \phi_\alpha\ran$ state when $|\alpha|^2\gtrsim N_s/e^2$.\

Let us study these two points separately.

\subsubsection{ \bf $|x_N|^2$ distribution} 


We first analyze the Glauber-state distribution, $| x_N|^2$, as a function of $N$ and $|\alpha|^2$. The $| x_N|^2$ distribution for elementary bosons, $|\alpha|^{2N}/N!$, has a maximum for $\ln (|\alpha |^2/N)\simeq 0$, that is, a peak at $N\simeq |\alpha |^2$. By contrast,  the $| x_{N}|^2$ distribution for Frenkel-like cobosons has two extrema that occur for $\ln \big[|\alpha |^2/N(1-N/N_s)\big]\simeq 0$, that is, at $N=N_\pm^*$ with  
\be 
N_\pm^*\simeq \frac{N_s}{2}\Big(1\pm \sqrt{1-4|\alpha |^2/N_s}\Big)
\ee
provided that $|\alpha |^2<N_s/4$.\

When $|\alpha |^2 \ll N_s$, the $| x_N|^2$ distribution has a pronounced peak at $N_-^*\simeq |\alpha |^2$ and a shallow minimum at $N_+^*$ very close to $N_s$, which is the maximum number of cobosons the sample can accommodate. This distribution looks very much like the one for elementary bosons.\

\begin{figure}[t!]
\begin{center}
\includegraphics[trim=3.5cm 6.5cm 6cm 2cm,clip,width=3.3in] {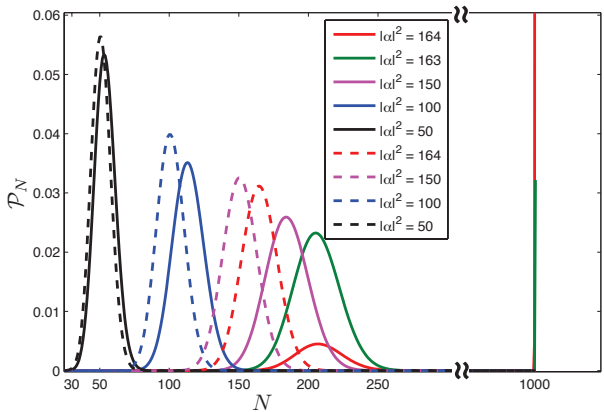}
   \caption{\small (color online) Occupation probability $\mathcal{P}_N$ for Frenkel-like cobosons, defined in Eq.~(\ref{eq:PNdistri}), as a function of $N$ for  different $|\alpha|^2$ when $N_s=1000$ (solid curves). For $N=N_s$, the green solid curve and the red solid curve go up to $0.03$ and $0.81$, respectively. The dashed curves correspond to elementary bosons.    }
   \label{Fig3}
\end{center}
\end{figure}

The  fermionic nature of the $\hat B^\dag$ coboson  constituents shows up when $ |\alpha |^2 $ scales as $N_s$. For $ |\alpha |^2=N_s/4$, the maximum of the $| x_N|^2$ distribution coincides with its minimum, $N^*_-=N^*_+$, while the distribution maximum disappears for larger $ |\alpha |^2$. The  Pauli exclusion principle then acts on the Fock states $|\hat N\ran$ in the most dramatic way. The number-operator eigenvalue of large-$N$ Fock states $|\hat N\ran$ is significantly reduced, as seen from Fig.~\ref{Fig1}. On the other hand, $N$ cannot be larger than $N_s$. So, in order for the quasi-Glauber states $|\hat \phi_\alpha\ran$ to produce an average boson number of the order of $N_s$,  these large-$N$ Fock states $|\hat N\ran$ must have a $| x_N|^2$ prefactor large enough to compensate for small number-operator eigenvalues. \

To quantify the consequences of Pauli blocking on Glauber states, we first note that 
\be
| x_{N_s-N}|^2|x_{N}|^2=| x_{N_s}|^2\, .
\ee
This relation allows us to write the occupation probability $\mathcal{P}_N$ for the Fock state $|\hat N\ran$ in two ways,
\be
\mathcal{P}_N=\frac{| x_{N}|^2}{\sum_{N'=0}^{N_s} | x_{N'}|^2}=\frac{| x_{N_s-N}|^{-2}}{\sum_{N''=0}^{N_s} | x_{N''}|^{-2}}\, .\label{eq:PNdistri}
\ee
 Figure \ref{Fig3} shows this probability as a function of $N$ for various $|\alpha|^2$ when $N_s=1000$. For elementary bosons (dashed curves),  a pronounced peak exists at $N=|\alpha|^2$ whatever $\alpha$. For Frenkel-like cobosons, the $\mathcal{P}_N$ probability has a similar peak when $|\alpha|^2$ is small. With increasing $|\alpha|^2$,  this peak  decreases in magnitude, its position is shifted toward higher $N$, although the average boson number
\be \frac{\lan \hat \phi_\alpha| \hat B^\dag \hat B |\hat \phi_\alpha\ran}{ \lan \hat \phi_\alpha| \hat \phi_\alpha\ran}=  |\alpha |^2\left(1-  \frac{| x_{N_s}|^2}{ \lan \hat \phi_\alpha| \hat \phi_\alpha\ran}\right)=|\alpha|^2 (1-\mathcal{P}_{N_s})\, ,\label{averbonNglau}
\ee
stays close to $|\alpha|^2$, since $\mathcal{P}_{N_s}$ then is very small. When $|\alpha|^2$ reaches $\simeq 0.16 N_s$, the peak starts to strongly decrease  without spreading, its missing weight being transferred to $|\hat N_s\ran$ and its neighbor states, thereby causing the probability $\mathcal{P}_{N_s}$ to increase abruptly up to 1. This rapid increase of $\mathcal{P}_{N_s}$ as $|\alpha|^2$ passes $0.16N_s$ makes the average boson number in the $| \hat \phi_\alpha\ran$ state decrease down to 0.\

\begin{figure}[t!]
\begin{center}
\includegraphics[trim=2.6cm 5.8cm 1cm 2.5cm,clip,width=3.8in] {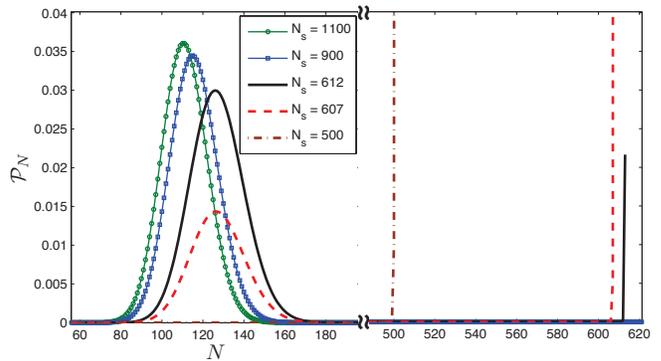}
   \caption{\small (color online) Occupation probability $\mathcal{P}_N$  for Frenkel-like cobosons, defined in Eq.~(\ref{eq:PNdistri}), as a function of $N$ for fixed $|\alpha|^2=100$ and different $N_s$. For $N=N_s$, the brown dot-dashed curve, the red dashed curve, and the black solid curve reach $0.99$, $0.52$, and $0.02$, respectively. }
   \label{Fig4}
\end{center}
\end{figure}

Figure \ref{Fig4} shows the $| x_{N}|^2$ distribution as a function of $N$ for fixed $|\alpha|^2$ and different $N_s$. The existence of a finite upper boundary, $N_s$, for the number of pair states forming the Frenkel-like coboson makes (i) the peak position shift toward high $N$ and (ii) the peak height decrease to its utter disappearance.


\subsubsection{\bf  $\hat g_2$ correlation function for the $|\hat \phi_\alpha\ran$ state}

We now consider the second-order correlation function $\hat g_2$ for the quasi-Glauber state $|\hat \phi_\alpha\ran$. As $x_{N_s}=\alpha  x_{N_s-1} $, we find, using Eq.~(\ref{eq:Bhatpsi_alpha}), 
\be
\hat B^2|\hat \phi_\alpha\ran=\alpha^2\sum_{N=0}^{N_s-2}  x_N |\hat N\ran\, .
\ee 
This gives $\hat g_2$ correlation function of the $ |\hat \phi_\alpha\ran$ state as
\be
\hat g_2=\frac{X_{N_s}X_{N_s-2}}{X_{N_s-1}^2}=1+\frac{| x_{N_s}|^2 X_{N_s-2}- | x_{N_s-1}|^2 X_{N_s-1}}{X_{N_s-1}^2}\label{eq:g2phialpha}
\ee
with $X_M=\sum_{N=0}^M | x_N|^2$.
As $| x_{N_s}|^2=|\alpha|^2 |x_{N_s-1}|^2$ while $|\alpha|^2| x_{N}|^2/| x_{N+1}|^2=(N+1) \hat F_{N+1}/\hat F_N$, we can rewrite the above equation as
\bea
\hat g_2-1&=&\frac{ |x_{N_s-1}|^2}{X_{N_s-1}^2} \big(|\alpha|^2 X_{N_s-2}-X_{N_s-1}\big) \label{eq:hatg2-1}\\
&=&\frac{ | x_{N_s-1}|^2}{X_{N_s-1}^2} \sum_{N=0}^{N_s-1}| x_{N}|^2\frac{(N_s-N)(N-1)}{N_s}\, .\nn
\eea
This shows that $\hat g_2$ is always larger than 1 since the unique negative contribution to the sum, which comes from the $N=0$ term, is small compared to the $N=2$ term for $|\alpha|^4> 2(N_s-1)/(N_s-2)$, a condition fulfilled for physically relevant $\alpha$'s.\

The fact that $\hat g_2$ for the quasi-Glauber state $|\hat \phi_\alpha\ran$  is larger than 1, while it is equal to 1 for elementary bosons, again comes from the Pauli exclusion principle:   Fock states for elementary bosons have a  $g_2$ correlation function equal to $1-1/N$; it always is smaller than 1 and approaches 1 when $N$ goes to infinity. By contrast,  coboson Fock states do not exist for $N$ larger than $N_s$ because of Pauli blocking; so, $N$ cannot increase up to infinity. Moreover, coboson Fock states have a second-order correlation function larger than 1 for $N>N_s/2+1$, again due to Pauli blocking. As a result, if large-$N$ Fock states $|\hat N\ran$ have a substantial weight in $|\hat \phi_\alpha\ran$, the resulting $\hat g_2$ also is larger than 1.
By contrast, if large-$N$ Fock states have a negligible weight in $|\hat \phi_\alpha\ran$, which occurs for $|\alpha|^2\ll N_s$, the $\hat g_2$ correlation function  goes to 1 because of the $| x_{N_s-1}|^2$ factor in Eq.~(\ref{eq:hatg2-1}). In such a case, the coboson composite nature plays a minor role.\
\begin{figure}[t!]
\begin{center}
\includegraphics[trim=0.5cm 0cm 0.5cm 0.7cm,clip,width=3.2in] {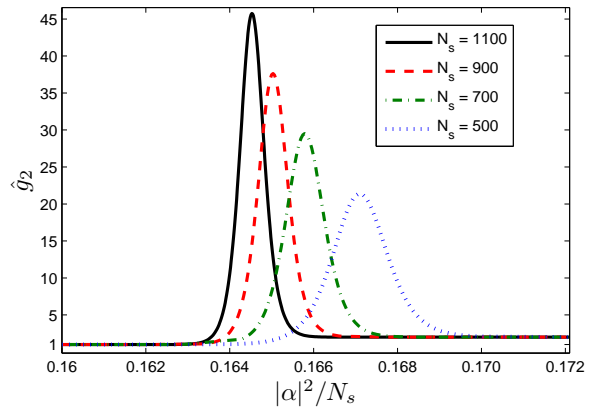}
   \caption{\small (color online) Second-order correlation function $\hat g_2$ as a function of $|\alpha|^2/N_s$ for the quasi-Glauber state $|\hat \phi_\alpha\ran$ of Frenkel-like cobosons, defined in Eq.~(\ref{eq:hatg2-1}). $\hat g_2$ stays equal to 1 on the left of the sharp peak that appears at $|\alpha|^2\simeq 0.16N_s$, but  stays equal to $2(1-1/N_s)$ on the right of this peak.     }
   \label{Fig5}
\end{center}
\end{figure}

Figure~\ref{Fig5} shows the second-order correlation function $\hat g_2$ for the quasi-Glauber state $|\hat \phi_\alpha\ran$ as a function of $|\alpha|^2$ for various $N_s$. At the threshold $|\alpha|^2\simeq 0.16 N_s$ for the disappearance of the $\mathcal{P}_N$ peak in Fig.~\ref{Fig3}, $\hat g_2$ exhibits a sharp peak, indicating a strong number fluctuation. For smaller $|\alpha|^2$, the $\hat g_2$ correlation function stays equal to 1, within an exponentially small correction, while for larger $|\alpha|^2$, it stays equal to $2(1-1/N_s)$. This value obtained for $|\alpha|^2$ larger than $\simeq 0.16N_s$ can be understood as follows: in this regime, the whole occupation probability $\mathcal{P}_N$ shifts toward the Fock state $|\hat N_s\ran$, as can be seen from Figs.~\ref{Fig3} and \ref{Fig4}, while the $\hat g_2$ correlation function for $|\hat N_s\ran$, as given in Eq.~(\ref{g2frenkelexciton}), is equal to $2(1-1/N_s)$.\

We conclude that, in the case of cobosons having a $\varphi_p$ distribution equal to a phase, as for Frenkel excitons, it is possible to wash out the quantum nature of the coboson field for $|\alpha|^2$ smaller than a threshold value $\simeq 0.16N_s$, above which the Pauli exclusion principle between the coboson fermionic constituents completely destroys its coherent nature, the quasi-Glauber state being essentially reduced to its $|\hat N_s\ran$ component.

\section{Composite bosons with a localized distribution\label{ssec4}}

We have shown that, for a coboson field $\hat B^\dag$ having a flat $|\varphi_p|$ distribution, the number of fermion-pair states making the coboson must be finite in order to have a normalized operator. As a direct consequence, it is not possible to construct exact Glauber states, i.e., eigenstates of the destruction operator $\hat B$. By contrast, eigenstates of the number operator $\hat B^\dag\hat B$ do exist but their eigenvalue is smaller than that for elementary bosons --- a signature of Pauli blocking on $N$-coboson states. \

We now consider cobosons $B^\dag=\sum_p\varphi_pB_p^\dag$ having a localized $\varphi_p$ distribution with $\sum_p |\varphi_p|^2=1$ fulfilled for an infinite number of free-pair states $p$, as in the case of Wannier excitons and most composite bosons which have a relative-motion wave function localized in space. We are going to show that exact eigenstates of the number operator and the destruction operator of the Wannier-like coboson field do exist. As a result, the bosonic nature of the field and the fermionic nature of its constituents can be completely erased. Because the construction of these states is quite complicated, we  relegate most derivations in \ref{app:sec3}, and shall here focus  on their physical steps.

\subsection{Fock states}

{\bf (1)} A first idea for the number-operator eigenstates is to take all cobosons in the same state, namely $B_N^{\dag N}|0\ran$ with
\be
B^\dag_N=\sum_p g_{N,p} B^\dag_p\, ,\label{BdagNdag}
\ee
the prefactors possibly depending on $N$. However, as shown in \ref{app:sec3a}, enforcing $(B^\dag B-\mathcal{N}_N)B^{\dag N}_N|0\ran=0$ leads to $|\varphi_{p}|^2$ independent of $p$, that is, the flat distribution previously considered.

{\bf (2)} Another idea for the number-operator eigenstates is to still take  a product of coboson operators, but with operators all different, namely $B^\dag_1 B^\dag_2\cdots B^\dag_N|0\ran$, with $B^\dag_N$ defined in Eq.~(\ref{BdagNdag}). As shown in \ref{app:sec3a} for $N=2$, we can write ($g_{1,p}$, $g_{2,p}$) in terms of the $\varphi_{p}$ distribution of the $B^\dag$ field, but with $g_{1,p}\not=g_{2,p}$. The number eigenvalue $\mathcal{N}_2$ follows from  
\be
1=\sum_p \frac{|\varphi_{p}|^2}{2|\varphi_{p}|^2+ \mathcal{N}_2 -1} \, .
\ee 
Here again, as $1=\sum_p |\varphi_{p}|^2$, Pauli blocking reduces the number eigenvalue $\mathcal{N}_2$ to less than 2.\

The procedure for $N=2$ can be  extended to higher $N$ in a straightforward manner, with similar though more complicated eigenstates. We thus conclude that Fock states can be constructed for a $B^\dag$ coboson field having a localized distribution, these Fock states being products of \textit{different} coboson operators.

\subsection{Glauber states}

A first idea, which actually works, is to look for eigenstates of the destruction operator $B$ as 
 \be
 |\tilde\phi_\alpha\ran=\sum_{N=0}^\infty (B^\dag_N)^N|0\ran\label{GCstateWannier}
 \ee
where $B^\dag_N$, defined in Eq.~(\ref{BdagNdag}), now depends not only on $N$ but also on $\alpha$. To fulfill  $B|\tilde\phi_\alpha\ran=\alpha |\tilde\phi_\alpha\ran$ amounts to fulfilling $B B_N^{\dag N}|0\ran=\alpha B_{N-1}^{\dag N-1}|0\ran$ in each $(N-1)$-coboson subspace. To understand how the solution of this equation develops, let us  consider the first few $N$'s.\

$\bullet$ For $N=1$, the condition $BB_1^\dag|0\ran=\alpha|0\ran$ is fulfilled for $g_{1,p}=\alpha \varphi_{p}$.\

$\bullet$ For $N=2$, the condition $BB_2^{\dag 2}|0\ran=\alpha B_1^\dag|0\ran$ is fulfilled for 
 \be
g_{2,p}=\frac{S_2- \sqrt{S_2^2-2\alpha \varphi_{p}^* g_{1,p}}}{2\varphi_{p}^* }\, ,\label{def:g2p}
\ee 
 with $S_N=\sum_p  \varphi_{p}^* g_{N,p}$. This gives $S_2$ through
\be
2 =\sum_p \left(1-\sqrt{1-\frac{2\alpha}{S_2^2} \varphi_{p}^* g_{1,p} }\right)\, .
\ee
As $|\varphi_{p}|^2$ scales inversely with sample volume, the $|\varphi_{p}|^2$ expansion of the above two equations gives
\bea
S_2&=& \frac{\alpha}{\sqrt{2!}}\left(1+\frac{\tau_1}{2}+\cdots\right)\, ,\\
g_{2,p}&=&\frac{\alpha^2\varphi_p}{2!S_2}\Big(1+|\varphi_{p}|^2+\cdots \Big)\, ,
\eea
where $\tau_n$ defined as
\be
\tau_n=\sum_p |\varphi^2_p|^{n+1}\label{eq:taun}
\ee 
physically corresponds to fermion exchange between $(n+1)$ cobosons\cite{Moniquebook}.\

$\bullet$ For $N=3$, the condition $BB_3^{\dag 3}|0\ran=\alpha B_2^{\dag 2}|0\ran$ is fulfilled for 
 \be
g_{3,p}=\frac{S_3- \sqrt{S_3^2-\left(\frac{8S_2}{3S_3}\right)\alpha \varphi^*_{p} g_{2,p} }}{4\varphi_{p}^* }\, ,\label{g3ps3}
\ee 
which gives $S_3$  through
\be
4=\sum_p \left(1-\sqrt{1-\left(\frac{8S_2}{3S^3_3}\right)\alpha \varphi^*_{p} g_{2,p}}\right)\, .\label{4sump1-1}
\ee
The resulting $S_3$ and $g_{3,p}$ expanded in $|\varphi_{p}|^2$ are given in \ref{app:sec3b}.

$\bullet$ For arbitrary-$N$ cobosons, a similar procedure gives, according to \ref{app:sec3b},
\be
g_{N,p}=\frac{S_N- \sqrt{S_N^2-4\alpha \frac{N-1}{N} \varphi_{p}^* g_{N-1,p}\left(\frac{S_{N-1}}{S_N}\right)^{N-2}}}{2(N-1)\varphi_{p}^* }\, .\label{eq:gpmNP22}
\ee 
The deduced $S_N$ expands as
\be
S_N=\frac{\alpha}{(N!)^{1/N}}\left(1+\frac{N-1}{2}\tau_1+\cdots\right)\, ,\label{sNexpansion22}
\ee
from which we get the first terms of the $g_{N,p}$ expansion in $|\varphi_{p}|^2$ as
\be
g_{N,p}=\frac{\alpha^N\varphi_p}{N!S_N^{N-1}}\left(1+\frac{N(N-1)}{2}|\varphi_{p}|^2+\cdots\right)\, .\label{gNpexpansion22}
\ee

Inserting these $g_{N,p}$'s  into Eq.~(\ref{GCstateWannier}), we obtain the Glauber state $|\tilde\phi_\alpha\ran$ for Wannier-like cobosons. It depends on the $\varphi_p$ distribution in a complicated manner; but through it, the bosonic nature of the Wannier-like coboson field and the fermionic nature of its constituents is completely erased, which is a formidable challenge!

 \section{Comparison with the Anderson's ansatz\label{ssec:6}}
  
Let us now compare the Glauber states we have constructed for Frenkel-like and Wannier-like cobosons with the ansatz proposed by Anderson for BCS condensate\cite{Anderson1958} but which has also been used for the exciton Bose-Einstein condensate\cite{Keldysh1968,Comte1982}.\
                                                                                
Anderson has shown\cite{Anderson1958} that for $B^\dag$ creating a coboson made of two electrons with opposite spins and opposite momenta, $B^\dag=\sum_\vp \varphi_\vp a^\dag_{\vp,\uparrow} a^\dag_{-\vp,\downarrow}$, the state
\be
|\Phi_\alpha\ran=\sum_{N=0}^\infty \frac{\alpha^N}{N!}   B^{\dag N}|0\ran=e^{\alpha  B^\dag}|0\ran\label{genAnderson}
\ee
reduces to the BCS state\cite{Moniquebook} 
\be
\prod_\vp \Big(u_\vp+v_\vp a^\dag_{\vp,\uparrow} a^\dag_{-\vp,\downarrow}\Big)|0\ran\label{BCS}
\ee
for $\alpha\varphi_\vp=v_\vp/u_\vp$. Pauli blocking acts on the $N$-coboson state in Eq.~(\ref{genAnderson}) through the fact that $(a^\dag_{\vp,\uparrow} a^\dag_{-\vp,\downarrow})^n=0$ for $n\geq 2$.  It is worth noting that for $B^\dag$ replaced by the elementary boson creation operator $A^\dag$, Eq.~(\ref{genAnderson}) reduces to the elementary boson Glauber state. So, the above ansatz must possess some kind of coherent character. Before going further, we wish to stress that $|\Phi_\alpha\ran$ is not eigenstate of the $B$ destruction operator, and therefore it is not an exact Glauber state with full coherence at all orders.
 
 \subsection{Frenkel-like cobosons}
 
  To study the probability distribution of the $|\Phi_\alpha\ran$ ansatz with respect to $N$ and $|\alpha|^2$ in the case of Frenkel-like cobosons, we replace $B^\dag$ with $\hat B^\dag$ given in Eq.~(\ref{eq:Bdagthetap}). 
By using the same algebraic manipulations that transform Eq.~(\ref{genAnderson}) into Eq.~(\ref{BCS}), we find that the Anderson's ansatz reads as
\be
|\hat\Phi_\alpha\ran=\prod_p \left(1+ \frac{\alpha}{\sqrt{N_s}} e^{i\theta_p}\alpha_p^\dag \beta_p^\dag\right)|0\ran\, .
\ee
As its normalization factor is given by
\be
\lan\hat \Phi_\alpha|\hat \Phi_\alpha\ran=\prod_p \left(1+ \frac{|\alpha|^2}{N_s} \right)=\left(1+ \frac{|\alpha|^2}{N_s} \right)^{N_s}\, ,
\ee
we find the Fock-state probability  in the $|\hat\Phi_\alpha\ran$ state as  
\be
\hat{\mathcal{P}}_N =\frac{|\lan \hat N|\hat\Phi_\alpha \ran|^2}{\lan\hat \Phi_\alpha|\hat \Phi_\alpha\ran}=\frac{|\alpha|^{2N}\hat F_N}{N!\left(1+ |\alpha|^2/N_s \right)^{N_s}}\, .\label{PNAnderson}
\ee
This probability exactly corresponds to the one obtained by Kaplan and Ruvinskii\cite{Kaplan1976}; so, the state they considered (see Eq.~(55) of Ref.~\onlinecite{Kaplan1976}) is not a true Glauber state. The $\hat{\mathcal{P}}_N $ distribution has a maximum at $N^*=|\alpha|^2/(1+|\alpha|^2/N_s)$ whatever $\alpha$. This $N^*$ number, which always is smaller than $N_s$, also is the average boson number of the $|\hat \Phi_\alpha\ran$ state.
\begin{figure}[t!]
\begin{center}
\includegraphics[trim=0cm 0cm 1cm 0cm,clip,width=3.2in] {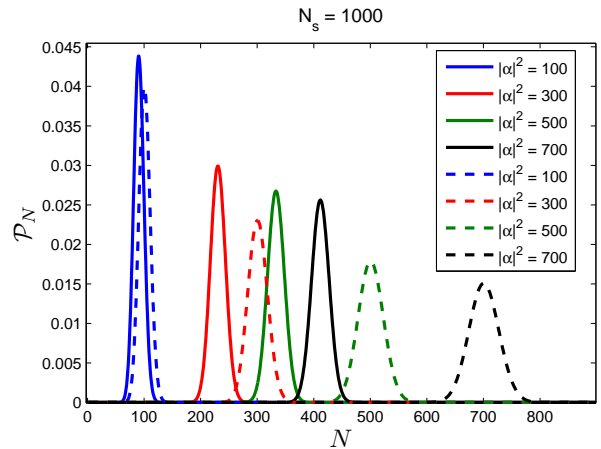}
   \caption{\small (color online) Occupation probability $\hat{\mathcal{P}}_N $ for the Anderson's ansatz defined in Eq.~(\ref{PNAnderson}), as a function of $N$ for different $|\alpha|^2$ (solid curves) when $N_s=1000$. The corresponding probability for elementary bosons is given for comparison (dashed curves). }
   \label{Fig6}
\end{center}
\end{figure}

 Figure~\ref{Fig6} shows the probability distribution of the $|\hat\Phi_\alpha\ran$ state as a function of $N$ for different $|\alpha|^2$. By comparison, Frenkel-like cobosons have a sharper peak. The larger the $|\alpha|^2$ value gets, the further their peak position shifts from $|\alpha|^2$ toward lower $N$. The peak intensity and the position shifting stand in stark contrast to those of the quasi-Glauber state $|\hat \phi_\alpha\ran$ given in Eq.~(\ref{eq:hatphialpha01}). \

Difference between the Glauber state $|\hat \phi_\alpha\ran$ and the Anderson's ansatz  $|\hat\Phi_\alpha\ran$  for Frenkel-like cobosons can also be seen from their second-order correlation function: $\hat g_2$ for $|\hat\Phi_\alpha\ran$ exactly reads 
\be
\hat g_2=1-\frac{1}{N_s}\, ,
\ee
whatever $\alpha$. By contrast, $\hat g_2$ for $|\hat \phi_\alpha\ran$  stays equal to 1 for $|\alpha|^2$ smaller than $\simeq 0.16N_s$, while for larger $|\alpha|^2$ it is equal to $2(1-1/N_s)$, which is twice the $\hat g_2$ value for the $|\hat\Phi_\alpha\ran$ state.\

We conclude that in the case of Frenkel-like cobosons, the Anderson's ansatz  $|\hat\Phi_\alpha\ran$ and the quasi-Glauber state $|\hat \phi_\alpha\ran$  differ qualitatively at large particle density. The $|\hat\Phi_\alpha\ran$ state, by construction, has a $\hat g_2$ less than 1; therefore, a residue of its bosonic quantum nature remains, whatever $\alpha$. 

   \subsection{Wannier-like cobosons}

Using Eqs.~(\ref{sNexpansion22}) and (\ref{gNpexpansion22}), it is easy to show that the $N$-coboson component $(\alpha^N/N!) B^{\dag N}|0\ran$ of the Anderson's ansatz $|\Phi_\alpha\ran$ corresponds to the leading expansion term of the $N$-coboson component of the Glauber state $|\tilde\phi_\alpha\ran$ given in Eq.~(\ref{GCstateWannier}). Difference results from Pauli blocking between the coboson fermionic constituents. As for Frenkel-like cobosons, the Glauber state $|\tilde\phi_\alpha\ran$ for Wannier-like cobosons is far more complex and definitely differs from the Anderson's ansatz $|\Phi_\alpha\ran$.

\section{State of the art\label{sota}}

Glauber states for composite bosons have been previously tackled by Kaszlikowski's group. Yet, the procedure they used\cite{Leepra2013} is  inconsistent because they have neglected the component of $B|\psi_N\ran$ that is orthogonal to the $|\psi_{N-1}\ran$ state, while keeping $F_{N}/F_{N-1}$ ratios different from 1 as unique signature of the particle composite nature. In doing so, they missed the fact that cobosons for which the orthogonal component exactly cancels, do exist, such as Frenkel-like cobosons whose $\varphi_p$ distribution is just a phase.\

To possibly construct Glauber states from the Fock states of cobosons having a flat $|\varphi_p|$ distribution, it is crucial to recognize that the number of pair states making such cobosons must be finite. Consequently, because of Pauli blocking, it is not possible to pile up more Frenkel-like cobosons than the number of pair states at hand. This leads to very different Fock-state probability distribution in the resulting coboson Glauber state.


\section{Conclusion}

In this paper, we establish a formal procedure to construct Fock and Glauber states for composite bosons. We show that it is possible to erase the bosonic quantum nature of a coboson field and the fermionic quantum nature of its constituents, despite the known robustness of Pauli blocking. Yet, the Pauli exclusion principle leading to the ``moth-eaten effect", present in all coboson systems, shows up in different ways. \

\noindent $\bullet$ For cobosons whose $\varphi_p$ distribution in their Schmidt decomposition  is just a phase, such as Frenkel excitons, Pauli blocking imposes a maximum  number $N_s$ of cobosons which strongly affects the Fock-state probability distribution in the Glauber states. This probability exhibits a sharp peak which resembles that of elementary-boson Glauber states, not only in the very dilute limit $|\alpha|^2 \ll N_s$, but also up to the rather dense regime  $|\alpha|^2 \lesssim 0.16N_s$. In this regime, the second-order correlation function $\hat g_2$ is essentially equal to 1; thus, the quantum nature of the coboson field is essentially erased. Beyond $|\alpha|^2\simeq 0.16N_s$, the Glauber states collapse to its $|\hat N_s\ran$ component. In this regime, the Pauli exclusion principle between the coboson fermionic constituents is quite strong and cannot be erased. 

\noindent $\bullet$ For cobosons having a localized $\varphi_p$ distribution, such as Wannier excitons, exact Fock and Glauber states can  be constructed. These  two states are far more complex than the ones for a flat distribution, because of the structured momentum dependence of $\varphi_p$. For both states, the coboson distribution in the $N$-coboson Fock state must be adjusted to compensate for Pauli blocking, which becomes stronger and stronger when the coboson number  increases.  Nevertheless, their behaviors with respect to  Pauli blocking are qualitatively the same as for Frenkel-like cobosons.   \

The Glauber states for composite bosons we have here constructed should be of great value for studying the output light source emitted by a conglomerate of nano-scale quantum emitters\cite{Sung2015,HeNT2015}, or by an exciton or polariton condensate\cite{Imanoglu1996,JinPRA2004,KimPRX}. A first direct application is the superradiance of Frenkel-like cobosons at high excitation density, which should display interesting phenomena different from those at low density\cite{Tokihiro1993,JinPRB2003}. The imprint of Pauli blocking should appear in their emission or absorption spectrum when the density increases.

\section*{Acknowledgement}

 S.-Y.S. acknowledges a three-month financial support from CNRS (France) as invited researcher at INSP in Paris. We wish to thank Fran\c{c}ois Dubin for having brought our attention to this interesting problem.

\renewcommand{\thesection}{\mbox{Appendix~\Roman{section}}} 
\setcounter{section}{0}
\renewcommand{\theequation}{\mbox{A.\arabic{equation}}} 
\setcounter{equation}{0} %
\section{Elementary boson field \label{app:sec1}}

Elementary-boson creation operators $A^\dagger$ obey the commutation relations 

\be
\big[A^\dagger, A^\dagger\big]_- = 0,\qquad \big[A,A^\dagger\big]_- = 1\, .
\ee
The quantum property of $N$ elementary bosons follows from 
\bea
\big[A, A^{\dagger N}\big]_-&=& \big[A, A^\dagger\big]_-A^{\dagger N-1}+A^\dagger\big[A, A^{\dagger N-1}\big]_-\nn\\
&=& N A^{\dagger N-1}\,.
\eea
The above equation readily gives 

\be\label{eq:twoprime}
AA^{\dagger N} \vert 0 \rangle = N A^{\dagger N - 1} \vert 0 \rangle\,,
\ee
which leads to 
\be 
\lan 0| A^N A^{\dag N}|0\ran=N!\,.\label{Nf_elemnboson}
\ee  
The normalized states $\vert N \rangle = (N!)^{-1/2} A^{\dagger N} \vert 0 \rangle$ are thus related by
\be
A \vert N \rangle = \sqrt{N} \vert N - 1 \rangle\, ,\label{eq:numberstates}
\ee
from which we readily get
\be
A^\dag A |N\ran=N|N\ran\, .\label{Nstateeemboso}
\ee
So, $\vert N \rangle$ is eigenstate of the number operator $A^\dagger A$ with  eigenvalue $N$. The $|N\rangle$ states are usually called Fock or number states. \

The bosonic quantum nature of these Fock states is seen from the fact that their second-order correlation function differs from 1. Since $A^{\dagger 2}  A^2 \vert N \rangle = N (N - 1) \vert N \rangle$, we find that $g_2$ defined in Eq.~(\ref{eq:g2}) is equal to $1 - 1/N$ for $|\phi\ran=|N\ran$.

\renewcommand{\theequation}{\mbox{B.\arabic{equation}}} 
\setcounter{equation}{0} %
\section{Decomposition of $B|\psi_N\ran$ different from Eq.~(\ref{eq:BpsiN}) \label{app:sec2}}

In order to write $B|\psi_N\ran$ as a state along $|\psi_{N-1}\ran$ and a state orthogonal to $|\psi_{N-1}\ran$ in an easy way, we introduce the projector $P_\perp^{(N)}$ over the subspace orthogonal to $|\psi_N\ran$, defined in terms of the identity operator ${\rm I}_N$ in the $N$-pair subspace as
\be
{\rm I}_N=\frac{|\psi_N\ran\lan \psi_N|}{\lan \psi_N|\psi_N\ran}+P_\perp^{(N)}\, .
\ee 
So, $\big(P_\perp^{(N)}\big)^2=P_\perp^{(N)}$ and $P_\perp^{(N)}|\psi_N\ran=0$, as easy to check. By inserting ${\rm I}_{N-1}$ in front of the $(N-1)$-pair state $B|\psi_N\ran$, we get, with the help of Eq.~(\ref{pshiNnormalifact}), 
\bea
B|\psi_N\ran&=&|\psi_{N-1}\ran\frac{\lan \psi_{N}|\psi_N\ran}{\lan \psi_{N-1}|\psi_{N-1}\ran}+P_\perp^{(N-1)}B|\psi_N\ran\nn\\
&=& N\frac{F_N}{F_{N-1}}|\psi_{N-1}\ran+|R_{N-1}\ran\, ,\label{BpsiNRN-1}
\eea
where $|R_{N-1}\ran=P_\perp^{(N-1)}B|\psi_N\ran$ by construction is orthogonal to $|\psi_{N-1}\ran$.\

To get the normalization factor of the $|R_N\ran$ state, we first note, using Eq.~(\ref{eq:BpsiN}), that this state also reads
 \be
 |R_N\ran=(N+1)\left(1-\frac{F_{N+1}}{F_{N}}\right)|\psi_N\ran-N(N+1) C^\dag |\psi_{N-1}\ran\, .
 \ee
So, from the above two equations and 
\be
\lan\psi_N| C^\dag |\psi_{N-1}\ran=(N-1)!\big(F_N-F_{N+1}\big)\, ,
\ee
as obtained by projecting Eq.~(\ref{eq:BpsiN}) over $|\psi_N\ran$, we get the $|R_N\ran$ normalization factor as
\be
\lan R_N|R_N\ran=\!\left(1{+}N\frac{F_{N+2}}{F_{N+1}}{-}(N{+}1)\frac{F_{N+1}}{F_{N}} \right)\!(N+1)! F_{N+1},\label{normalfactorRN}
\ee
in agreement with Ref.~\onlinecite{Lawpra2005}. So, $\lan R_N|R_N\ran$ would reduce to zero for $F_N=1$, as in the case of elementary bosons. However, it is inconsistent to neglect this term because $F_{N+1}/F_N$ for cobosons expands as\cite{CombescotEPL}
\bea
\frac{F_{N+1}}{F_N}&=&1-N\tau_1+N(N-1)(\tau_2-\tau_1^2)\frac{F_{N-2}}{F_N}\label{eq:FNgeneral} \\
&&-N(N-1)(N-2)(\tau_3-\tau_2\tau_1) \frac{F_{N-3}}{F_N}+\cdots\nn
\eea
where $\tau_n=\sum_p |\varphi^2_p|^{n+1}$ comes from fermion exchange between $(n+1)$ cobosons\cite{Moniquebook}. So, $|R_N\ran$ differs from zero when $F_{N+1}/F_{N}$ differs from 1. \

We wish to note that $\lan R_N|R_N\ran$ cancels not only for $F_{N+1}/F_{N}=1$, but also for 
\be
\frac{F_{N+1}}{F_{N}} =1-N\tau_1\, ,\label{ratioFN1FN}
\ee 
which corresponding to $F_{N+1}/F_{N}$ in the dilute limit\cite{Combescotepjb2003}. For Frenkel-like cobosons having a  flat distribution $|\varphi_p|=1/\sqrt{N_s}$, the $\tau_n$ factor is equal to $1/N_s^n$; so, $F_{N+1}/F_{N}$ then is exactly equal  to the first two terms of Eq.~(\ref{eq:FNgeneral}). \

The above $1-N\tau_1$ value for $F_{N+1}/F_{N}$ actually is the lowest value that can be derived using the so-called Schmidt number $N^*_{eff}=1/\tau_1$ in quantum information theory\cite{Lawpra2005,Chudzicki2010}. This Schmidt number is commonly used to gauge quantum entanglement between two fermions in a coboson.

\renewcommand{\theequation}{\mbox{C.\arabic{equation}}} 
\setcounter{equation}{0} %
\section{Fock and Glauber states for a localized coboson distribution\label{app:sec3}} 
 
\subsection{Fock states\label{app:sec3a}}


{\bf (1)} A calculation similar to the one leading to Eq.~(\ref{eq:BpsiN}) gives
\bea
B B_N^{\dag N}|0\ran&=& N\Big(\sum_p \varphi_p^* g_{N,p}\Big)B^{\dag N-1}_N |0\ran\label{eq:B|BN^N0}\\
&&-N(N-1) \Big(\sum_p \varphi_p^* g^2_{N,p}B^\dag_p\Big)B^{\dag N-2}_N |0\ran\, ,\nn
\eea
 with $B^\dag_N$ defined in Eq.~(\ref{BdagNdag}). Enforcing $B_N^{\dag N}|0\ran$ to be eigenstate of the $B^\dag B$ operator, namely $(B^\dag B-\mathcal{N}_N)B^{\dag N}_N|0\ran=0$, leads to
\bea
0&=& \sum_{p_1 \cdots p_N}\Bigg\{\varphi_{p_1}\bigg( N\Big(\sum_p \varphi_p^* g_{N,p}\Big)g_{N,p_2}\cdots g_{N,p_N}\nn\\
&&-N(N-1)\varphi_{p_2}^* g^2_{N,p_2} g_{N,p_3}\cdots g_{N,p_N}\bigg)\nn\\
&&-\mathcal{N}_N g_{N,p_1}\cdots g_{N,p_N}\Bigg\} B^\dag_{p_1}B^\dag_{p_2}\cdots B^\dag_{p_N}|0\ran\, .\label{g_Npcondition}
\eea

A way to fulfill this condition is to set the curly bracket of the above equation to zero. For 
\be
S_N\equiv \sum_p  \varphi_p^* g_{N,p}\, ,
\ee
we then get
\be
0=\varphi_{p_1}\Big(NS_N-N(N-1)\varphi_{p_2}^* g^2_{N,p_2} \Big)-\mathcal{N}_N g_{N,p_1}
\ee
whatever $(p_1,p_2)$. The above equation also reads
\be
\frac{\mathcal{N}_N}{N}\frac{g_{N,p_1}}{\varphi_{p_1} }=K=S_N-(N-1) \varphi_{p_2}^* g_{N,p_2}\, ,\label{eq:NnNgnp1}
\ee
where $K$ does not depend on $(p_1,p_2)$. So,
\be
\varphi^*_{p}\varphi_{p}=\frac{S_N-K}{K}\frac{\mathcal{N}_N}{N(N-1)}\, .
\ee
As a result, $|\varphi_{p}|$ does not depend on $p$: this corresponds to the previously-studied flat distribution.\

Equation (\ref{g_Npcondition}) is for sure fulfilled by canceling its curly bracket, but this is a priori not the only way. Indeed, for  fermion pair operators $B^\dag_p=\alpha_p^\dag \beta_p^\dag$, we do have
\bea
\big[B_k,B^\dag_p\big]_-\!\!&=&\!\! \delta_{kp}(1{-}\alpha_p^\dag\alpha_p{-}\beta^\dag_p\beta_p){=}\delta_{kp}{-}D_{k,p}\, ,\\
\big[D_{k,p},B^\dag_{p'}\big]_-\!\!&=&\!\! 2\delta_{kp}\delta_{pp'} B^\dag_{p'}\, .
\eea
So, by projecting Eq.~(\ref{g_Npcondition}) taken for $N=2$ over the state $\lan 0| B_{k_2}B_{k_1}$ with $k_1\not=k_2$, we get
\be
\mathcal{N}_N \frac{g_{2,k_1}}{\varphi_{k_1}}\frac{g_{2,k_2}}{\varphi_{k_2}}=\frac{g_{2,k_1}}{\varphi_{k_1}}(S_2-\varphi^*_{k_1}g_{2,k_1})+\frac{g_{2,k_2}}{\varphi_{k_2}}(S_2-\varphi^*_{k_2}g_{2,k_2})\, .
\ee
Such a relation, of the form $G(x)G(y)=F(x)+F(y)$, imposes $G(x)$ and $F(x)$ to be $x$-independent, which leads to the same result as Eq.~(\ref{eq:NnNgnp1}). Calculation for $N>2$ yields the same conclusion.\



{\bf (2)} We now look for $(B^\dag B-\mathcal{N}_2)B^\dag_1 B^\dag_2|0\ran=0$. This equation reads
\bea
0&=&\sum_{p_1 p_2}\bigg\{\varphi_{p_1}\Big(g_{2,p_2} S_1+g_{1,p_2} S_2- 2\varphi^*_{p_2} g_{1,p_2}g_{2,p_2} \Big)\nn\\
&&-\mathcal{N}_2  g_{1,p_2}g_{2,p_1} \bigg\}   B^\dag_{p_1}B^\dag_{p_2}|0\ran\, .
\eea
By again setting the curly bracket to zero, we get 
\be
\mathcal{N}_2 \frac{g_{2,p_1}}{\varphi_{p_1}}=K'= \frac{g_{2,p_2}}{g_{1,p_2}}S_1+S_2-2\varphi^*_{p_2}g_{2,p_2}\, ,
\ee
which leads to 
\be
g_{2,p}=\frac{K'}{\mathcal{N}_2}\varphi_{p}\,, \quad  g_{1,p}=\frac{ S_1}{2|\varphi_{p}|^2+ \mathcal{N}_2 (1-S_2/K')}\varphi_{p}\, .
\ee
The first equation gives $S_2=K'/\mathcal{N}_2$. Inserting this result into the second equation, we end up with  
\be
1=\sum_p \frac{|\varphi_{p}|^2}{2|\varphi_{p}|^2+ \mathcal{N}_2 -1} \, ,
\ee 
which can be numerically solved for $\mathcal{N}_2$.



\subsection{Glauber states \label{app:sec3b}}

$\bullet$ For $N=3$, the expansions of Eqs.~(\ref{g3ps3}) and (\ref{4sump1-1}) in $|\varphi_{p}|^2$ read 
\bea
S_3&=&\frac{\alpha}{(3!)^{1/3}}\left(1+\tau_1+\cdots\right)\, ,\\
g_{3,p}&=&\frac{\alpha}{3}\left(\frac{S_2}{S_3^2}g_{2,p}+\alpha\frac{2 S_2^2}{3S_3^5}\varphi_p^*g_{2,p}^2+\cdots\right)\nn\\
&=&\frac{\alpha^3\varphi_p}{3!S_3^2}\left(1+3|\varphi_{p}|^2+\cdots\right)\, .
\eea
Again, we find that both $g_{3,p}$ and $S_3$ are linear in $\alpha$.\

$\bullet$ For arbitrary-$N$ cobosons, $B B_N^{\dag N}|0\ran$ calculated through Eq.~(\ref{eq:B|BN^N0}) yields
\be
0=\sum_{p_1\cdots p_N} G_{p_1\cdots p_{N-1}}B^\dag_{p_1}\cdots B^\dag_{p_{N-1}}|0\ran\, ,\label{eq:GNPBBB}
\ee 
which is fulfilled for
 \bea
0= G_{p_1\cdots p_{N-1}}&\equiv&N\Big(\sum_p \varphi_p^* g_{N,p}\Big)g_{N,p_1}\cdots g_{N,p_{N-1}}\nn\\
 &&-N(N-1)\varphi_{p_1}^* g^2_{N,p_1} g_{N,p_2}\cdots g_{N,p_{N-1}}\nn\\
 &&-\alpha g_{N-1,p_1} \cdots g_{N-1,p_{N-1}}\, .
 \eea
 If we multiply the above equation by $\varphi_{p_2}^* \varphi_{p_3}^*\cdots \varphi_{p_{N-1}}^* $ and then sum over $(p_2,\cdots,p_{N-1})$, we get
\be
0=S_N g_{N,p_1}-(N-1)\varphi_{p_1}^* g^2_{N,p_1} -\frac{\alpha}{N}g_{N-1,p_1}\left(\frac{S_{N-1}}{S_N}\right)^{N-2}\, .
\ee
So, $g_{N,p}$ reads in terms of $g_{N-1,p}$ as
\be
g^{(\pm)}_{N,p}=\frac{S_N\pm \sqrt{S_N^2-4\alpha \frac{N-1}{N} \varphi_{p}^* g_{N-1,p}\left(\frac{S_{N-1}}{S_N}\right)^{N-2}}}{2(N-1)\varphi_{p}^* }\, .\label{eq:gpmNP}
\ee 
We rule out the plus-sign solution because the associated coboson distribution $g^{(+)}_{N,p}$ has a term in $1/\varphi_{p}^*$, which delocalizes the distribution. As the $g^{(-)}_{N,p}$ distribution converges for large $p$, this allows us to construct the $ B_N^{\dag N}|0\ran$ state from the $ B_{N-1}^{\dag N-1}|0\ran$ state. The above equation expanded in $|\varphi_{p}|^2$ leads to
\bea
S_N\!\!&=&\!\!\frac{\alpha}{(N!)^{1/N}}\left(1+\frac{N-1}{2}\tau_1+\cdots\right)\, ,\label{sNexpansion}\\
g_{N,p}\!\!&=&\!\!\frac{\alpha}{N}\!\!\left(\frac{S_{N-1}^{N-2}}{S_N^{N-1}}g_{N-1,p}{+}\alpha\frac{N-1}{N}\frac{S_{N-1}^{2N-4}}{S_N^{2N-1}}\varphi_{p}^* g^2_{N-1,p} {+}\cdots\!\!\right)\nn\\
&=&\frac{\alpha^N\varphi_p}{N!S_N^{N-1}}\left(1+\frac{N(N-1)}{2}|\varphi_{p}|^2+\cdots\right)\, ,\label{gNpexpansion}
\eea
which agree with the results obtained for $N=(1,2,3)$. We again see that both $S_N$ and $g_{N,p}$ are linear in $\alpha$.



\begin{thebibliography}{99}

\bibitem{HBT1} R. Hanbury Brown, and R. Q. Twiss, Nature {\bf177}, 27 (1956).
\bibitem{HBT2} R. Hanbury Brown, and R. Q. Twiss, Nature {\bf178}, 1046 (1956).





\bibitem{Glauber1963} R. J. Glauber, Phys. Rev. {\bf 131}, 2766 (1963). 
\bibitem{Yarivbook} A. Yariv, {\it Qauntum Electronics}, 3rd ed. Wiley (1989).
\bibitem{Loudon} R. Loudon, {\it The Quantum Theory of Light}, Oxford Univ. Press (2000).


\bibitem{Yasuda1997} M. Yasuda, and F. Shimizu, Phys. Rev. Lett. {\bf77}, 3090 (1996).
\bibitem{Hodgman2011} S. S. Hodgman, R. G. Dall, A. G. Manning, K. G. H. Baldwin, and A. G. Truscott, Science, {\bf331}, 1046 (2011).
\bibitem{Guarrera2011}V. Guarrera, P. W\"{u}rtz, A. Ewerbeck, A. Vogler, G. Barontini, and H. Ott, Phys. Rev. Lett. {\bf107}, 160403 (2011).

\bibitem{Manning2013} A. G. Manning, W. RuGway, S. S. Hodgman,  R. G. Dall, K. G. H. Baldwin, and A. G. Truscott, New J. Phys, {\bf15}, 013042 (2013).



\bibitem{Dall2013NP} R. G. Dall, A. G. Manning, S. S. Hodgman, W. RuGway, K. V. Kheruntsyan, and A. G. Truscott, Nat. Phys. {\bf9}, 341 (2013).


\bibitem{MoniquePRA2009} M. Combescot, F. Dubin, and M. A. Dupertuis, Phys. Rev. A {\bf80}, 013612 (2009).
\bibitem{PogosovJETP2010} W. V. Pogosov and M. Combescot, JETP Letters {\bf92}, 484 (2010).

\bibitem{MoniqPhysreport} M. Combescot, O. Betbeder-Matibet, and F. Dubin, Physics Reports {\bf 463}, 215 (2008).

\bibitem{Moniquebook} M. Combescot and S.-Y. Shiau, {\it Excitons and Cooper Pairs: two composite bosons in many-body physics}, Oxford Univ. Press (2015).

 \bibitem{Agranovichbook} V. M. Agranovich, {\it Excitations in Organic Solids}, Clarendon Press, Oxford (2008).

\bibitem{Anderson1958} P. W. Anderson, Phys. Rev. {\bf112}, 1900 (1958).


\bibitem{MoniquePRB2008} M. Combescot and W. V. Pogosov, Phys. Rev. B {\bf77}, 085206 (2008).

\bibitem{Lawpra2005} C. K. Law, Phys. Rev. A {\bf71}, 034306 (2005).

\bibitem{Leepra2013} S.-Y. Lee, J. Thompson, P. Kurzy\'{n}ski, A. Soeda, and D. Kaszlikowski, Phys. Rev. A {\bf88}, 063602 (2013).




\bibitem{Keldysh1968} L. V. Keldysh and A. N. Kozlov, Sov. Phys. JETP {\bf27}, 521 (1968).
\bibitem{Comte1982} C. Comte and P. Nozi\`{e}res, J. Physique {\bf43}, 1069 (1982).

\bibitem{Kaplan1976} I. G. Kaplan and M. A. Ruvinskii, Sov. Phys. JETP {\bf 44}, 1127 (1976).

\bibitem{Sung2015} J. Sung, P. Kim, B. Fimmel, F. W\"{u}rthner, and D. Kim, Nat. Commun. 6:8646 doi:10.1038/ncomms9646 (2015). 

\bibitem{HeNT2015} Y. N. He, G. Clark, J. R. Schaibley, Y. He, M. C. Chen, Y. J. Wei, X. Ding, Q. Zhang, W. Yao, X. Xu, C. Y. Lu, and J. W. Pan, Nature Nanotech. {\bf 10}, 497 (2015). 

\bibitem{Imanoglu1996} A. Imamoglu, R. J. Ram, S. Pau, and Y. Yamamoto, Phys. Rev. A {\bf53}, 4250 (1996).
\bibitem{JinPRA2004} G. R. Jin and W. M. Liu, Phys. Rev. A {\bf70}, 013803 (2004).
\bibitem{KimPRX} S. Kim, B. Zhang, Z. Wang, J. Fischer, S. Brodbeck, M. Kamp,
C. Schneider, S. H\"{o}fling, and Hui Deng, Phys. Rev. X {\bf6}, 011026 (2016).


\bibitem{Tokihiro1993} T. Tokihiro, Y. Manabe, and E. Hanamura, Phys. Rev. B {\bf47}, 2019 (1993).



\bibitem{JinPRB2003} G. R. Jin, P. Zhang, Y.X. Liu, and C. P. Sun, Phys. Rev. B {\bf68}, 134301 (2003).

\bibitem{CombescotEPL} M. Combescot, Europhys. Lett. {\bf96}, 60002 (2011). 

\bibitem{Combescotepjb2003} M. Combescot, X. Leyronas, and C. Tanguy, Eur. Phys. J. B {\bf31}, 17 (2003).

\bibitem{Chudzicki2010} C. Chudzicki, O. Oke, and W. K. Wootters, Phys. Rev. Lett. {\bf104}, 070402 (2010).


\end{thebibliography}
\end{document}